\documentclass[useAMS,usenatbib,a4paper]{mn2e}
\usepackage{times}
\usepackage{graphicx}
\usepackage{amssymb}
\title[A discy origin for a Perseus dE]{Dwarf galaxies in the Perseus Cluster: further evidence for a disc origin for dwarf ellipticals}
\author[S. J. Penny et al.]{Samantha~J.~Penny$^{1,2}$, Duncan A. Forbes$^{3}$, Kevin A. Pimbblet$^{4,1,2}$ \newauthor and David J.~E. Floyd$^{1,2}$\\
$^1$School of Physics, Monash University, Clayton, Victoria 3800, Australia\\ 
$^2$Monash Centre for Astrophysics, Monash University, Clayton, Victoria 3800, Australia\\
$^3$Centre for Astrophysics and Supercomputing, Swinburne University of Technology, Hawthorn, Victoria 3122, Australia\\
$^4$ Department of Physics and Mathematics, University of Hull, Cottingham Road, Kingston-upon-Hull, HU6 7RX, United Kingdom\\
}
\begin{document}

\date{Received 2014 July 07; in original form 2014 May 28}

\pagerange{\pageref{firstpage}--\pageref{lastpage}} \pubyear{2002}

\maketitle

\label{firstpage}

\begin{abstract}
We present the results of a Keck-\textsc{esi} spectroscopic study of six dwarf elliptical (dE) galaxies in the Perseus Cluster core, and confirm two dwarfs as cluster members for the first time. All six dEs follow the size-magnitude relation for dE/dSph galaxies. Central velocity dispersions are measured for three Perseus dwarfs in our sample, and all lie on the $\sigma$-luminosity relation for early-type, pressure supported systems. We furthermore examine SA~0426-002, a unique dE in our sample with a bar-like morphology surrounded by low-surface brightness wings/lobes ($\mu_{B} = 27$~mag~arcsec$^{-2}$). Given its morphology, velocity dispersion ($\sigma_{0} = 33.9 \pm 6.1 $~km~s$^{-1}$), velocity relative to the brightest cluster galaxy NGC~1275 (2711~km~s$^{-1}$), size ($R_{e} =2.1 \pm 0.10$~kpc), and S\'ersic index ($n= 1.2 \pm 0.02$), we hypothesise the dwarf has morphologically transformed from a low mass disc to dE via harassment. The low-surface brightness lobes can be explained as a ring feature, with the bar formation triggered by tidal interactions via speed encounters with Perseus Cluster members. Alongside spiral structure found in dEs in Fornax and Virgo, SA~0426-002 provides crucial evidence that a fraction of bright  dEs have a disc infall origin, and are not part of the primordial cluster population.
\end{abstract}


\section{Introduction}

It is becoming increasingly clear that dwarf elliptical (dE) galaxies are not a simple, homogeneous galaxy population. Once thought to be the original building blocks in the hierarchical model of galaxy assembly that ceased their star formation early in the history of the Universe, dEs are instead found to exhibit a wide range of ages, metallicities, and morphologies inconsistent with a single primordial formation scenario (e.g. \citealt{2008MNRAS.383..247P,2009ApJ...706L.124L}). 

An alternative formation scenario is that a fraction of bright dEs are formed through the morphological transformation of low mass, disc galaxies via processes such as harassment and ram pressure stripping as they infall into galaxy groups and clusters. Through harassment and interactions, the progenitor galaxy becomes morphologically transformed into a dE \citep{1996Natur.379..613M,1998ApJ...495..139M,2001ApJ...559..754M,2005MNRAS.364..607M,2012MNRAS.424.2401V}. As the disc infalls into the cluster, it is stripped of its interstellar medium, and is dynamically heated by high-speed interactions with other galaxies and the gravitational potential of the cluster. To compensate, the galaxy loses stars, and over time the disc galaxy can morphologically transform into a dE. Evidence for this disc galaxy origin has been found in the form of embedded structures in cluster dEs such as bars and spiral arms \citep{2000A&A...358..845J,2003AJ....126.1787G,2003A&A...400..119D,2006AJ....132..497L,2009A&A...501..429L}.  

During harassment, bar formation may be triggered via disc instabilities \citep{2005MNRAS.364..607M,2009A&A...494..891A}, and the galaxy will loose stars from its disc until only the bar remains. Examples of bar-like dwarfs that are likely the result of harassment are found in the Virgo Cluster, including VCC794, VCC1392, and VCC1567 \citep{2012ApJ...750..121G}. These galaxies are sufficiently evolved that they show no evidence for ongoing mass loss, with only their central bars remaining. Combined with observations of rotational, rather than pressure, supported dwarfs with rotation curves similar to those of star forming galaxies \citep{2011A&A...526A.114T, 2014MNRAS.439..284R}, evidence is mounting that a fraction of dEs originated as low mass disc galaxies, and not as part of the primordial cluster population.

In this letter, we extend the study of dE kinematics to the Perseus Cluster, one of the richest and most relaxed galaxy clusters in the nearby Universe. As part of a wider study of dE kinematics across a range of environmental density (Penny et al., in prep.), we present Keck-ESI and HST-ACS observations of a sample of nucleated dEs in the Perseus Cluster core ($D=70$~Mpc) to examine their scaling relations. Here, we examine the kinematic properties of three dwarf ellipticals in the cluster core: two normal dwarf ellipticals (CGW38 and CGW39), and one unusual dE:  SA~0426-002. We go on to establish morphological transformation via harassment as a possible formation scenario for SA~0426-002. First presented in Conselice, Gallagher \& Wyse (2002),  SA~0426-002 exhibits an unusual structure, with a strong-bar like morphology, no evidence for ongoing star formation, and low surface brightness ``wings'' or ``lobes'' surrounding the central bar.  Conselice et al. (2002) noted the dwarf is likely undergoing dynamical interaction, and merits further attention to uncover its nature and origin. 

We describe our observations in Section~\ref{sec:obs},  and present the results  in Section~\ref{sec:res}, including the size-magnitude relation in Section~\ref{sec:sg}.  We discuss these results and the possibility of ongoing tidal evolution in Section~\ref{sec:discuss}, with our conclusion in Section~\ref{sec:conc}.

\section{Observations}
\label{sec:obs}

\subsection{Keck ESI spectrscopy}

\begin{figure}
\includegraphics[width=0.48\textwidth]{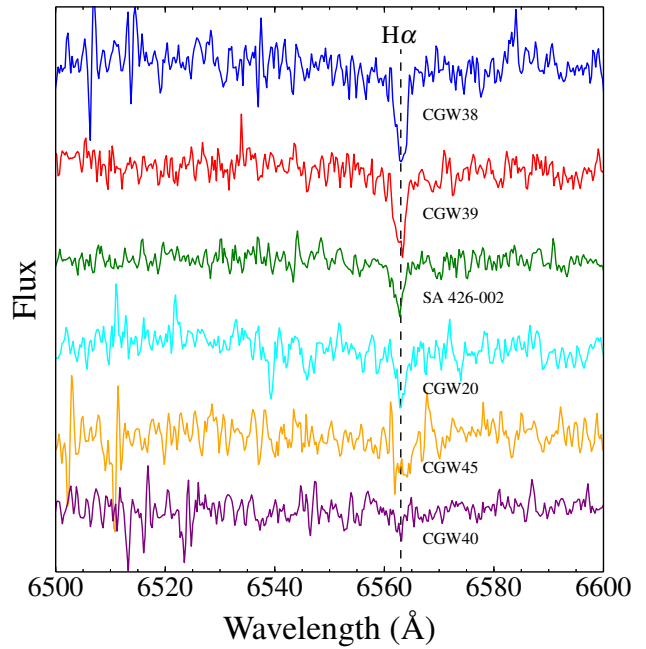}
\caption{Keck-\textsc{esi} spectra for the six Perseus dwarfs presented in this paper, trimmed to show the H$\alpha$ features (\textsc{esi} order 9). The spectra are corrected to rest wavelength using the velocities presented in Table~\ref{dwarfprops}. The spectra are sorted from top to bottom in order of decreasing S/N, and have not been smoothed. \label{fig:spec}}
\end{figure}

We observed bright dwarf ellipticals in the Perseus Cluster using the Echellette Spectrograph and Imager (\textsc{esi}) instrument \citep{2002PASP..114..851S}  on the Keck~\textsc{ii} telescope in echelle mode ($R \sim 8000$ for the $0.5''$ slit). Targets were selected from the catalogue of \citet{2003AJ....125...66C}, with their cluster membership confirmed spectroscopically by \citet{2008MNRAS.383..247P}, else through dE-like morphology in \textit{HST} imaging.  Observations were taken on the nights 2012 November 5th - 7th, with typical seeing 0.8''. We utilise the $0.5''$ slit, providing an instrument resolution $\sigma = 15.8$~km~s$^{-1}$.   The data provides complete wavelength coverage from 3900-10000~$\rm{\AA}$ over ten orders. The data were reduced using standard calibration techniques with the automated \textsc{makee} reduction package written by T. Barlow. Our spectroscopic observations are summarised in Table~\ref{tab:obs}, with spectra in the region of the H$\alpha$ feature for all six dwarfs shown in Figure~\ref{fig:spec}.  

Radial velocities for the target dEs were determined using the IRAF task \textsc{fxcor}, utilising the Fe/Mg (\textsc{esi} order 12), H$\alpha$ (\textsc{esi} order 9), and CaT (\textsc{esi} order 7) lines. We use standard stars taken under an identical observing setup as zero-redshift templates. We re-confirm cluster membership for four dwarfs with radial velocities measured by \citet{2008MNRAS.383..247P}, and confirm CGW40 and SA~0426-002 as members of the Perseus Cluster for the first time.

\begin{table}
\caption{Observing Parameters.\label{tab:obs}}
\begin{tabular}{lcccc}
\hline
Galaxy & S/N & Seeing & Exp. time \\
       &   &  ($^{''}$) &  (min) \\
\hline
SA~0426-002 & 7 & 0.8 & 80 \\
CGW~20 & 5 & 0.7 & 120 \\
CGW~38 & 10 & 0.6 & 90 \\
CGW~39 & 10 & 0.7 & 60 \\
CGW~40 & 5 & 0.5 & 150 \\
CGW~45 & 5 & 0.6 & 80 \\ 
\hline
\end{tabular}
\\
Notes: S/N is the minimum signal-to-noise in the continuum at $\sim$8500\AA. 
\end{table}

\begin{table*}
\caption{Properties of the Perseus dwarfs with spectroscopically confirmed cluster membership. The measured velocities and central velocity dispersions were determined from 
our Keck-ESI spectroscopy. The $B$-band magnitudes and sizes for the dwarfs are taken from Conselice, Gallagher \& Wyse (2003), and De Rijcke et al. (2009), apart from SA~0426-002, which has its size determined in Section~\ref{sec:sg}. Dynamical masses are calculated in Section~\ref{sec:veldisps}.}
\begin{center}
\begin{tabular}{lcccccccc}
\hline
Dwarf & $\alpha$ & $\delta$ & $M_{B}$ & v & $R_{e}$ & $\sigma_{0}$ & $M_{dyn}$ & $M_{\star}$\\
 & (J2000.0) & (J2000.0) & (mag) & (km~s$^{-1}$) & (kpc) & (km~s$^{-1}$) & ($M_\odot$) &  ($M_\odot$) \\
\hline
SA~0426-002 & 03:19:40.00 & +41:31:00.0 & -16.3 & $7987 \pm 15$ & 2.1 & $33.9 \pm 6.1$ & $(4.5\pm 2.3)\times10^{9}$ & $(1.9\pm0.8)\times10^{9}$\\
CGW~20 & 03:19:10.40 & +41:29:37.0 & -15.1 & $ 7325 \pm 16$ & 0.65  & \ldots & \ldots & \ldots \\
CGW~38  & 03:19:27.10 & +41:27:16.1 & -15.9 &$4205 \pm 23$ & 0.60 & $40.0 \pm 5.5$  &  $(1.8\pm 0.9)\times10^{9}$&$(1.9\pm0.3)\times10^{9}$ \\
CGW~39 & 03:19:31.40 & +41:26:28.7 & -15.7 & $6461 \pm 24$& 0.53 & $29.7 \pm 7.3$  &$(8.9 \pm 0.5)\times10^{8}$ & $(1.7\pm0.5)\times10^{9}$\\
CGW~40 & 03:19:31.70 & +41:31:21.3 & -15.1 & $3763\pm 15$ & 0.95 & \ldots &\ldots &\ldots \\
CGW~45 & 03:19:41.70 & +41:29:17.0 & -15.7 & $3156 \pm 25$ & 1.10 & \ldots & \ldots &\ldots \\
\hline
\end{tabular}
\end{center}
\label{dwarfprops}
\end{table*}%

\subsection{Hubble Space Telescope ACS imaging}

\begin{figure*}
\includegraphics[width=0.47\textwidth]{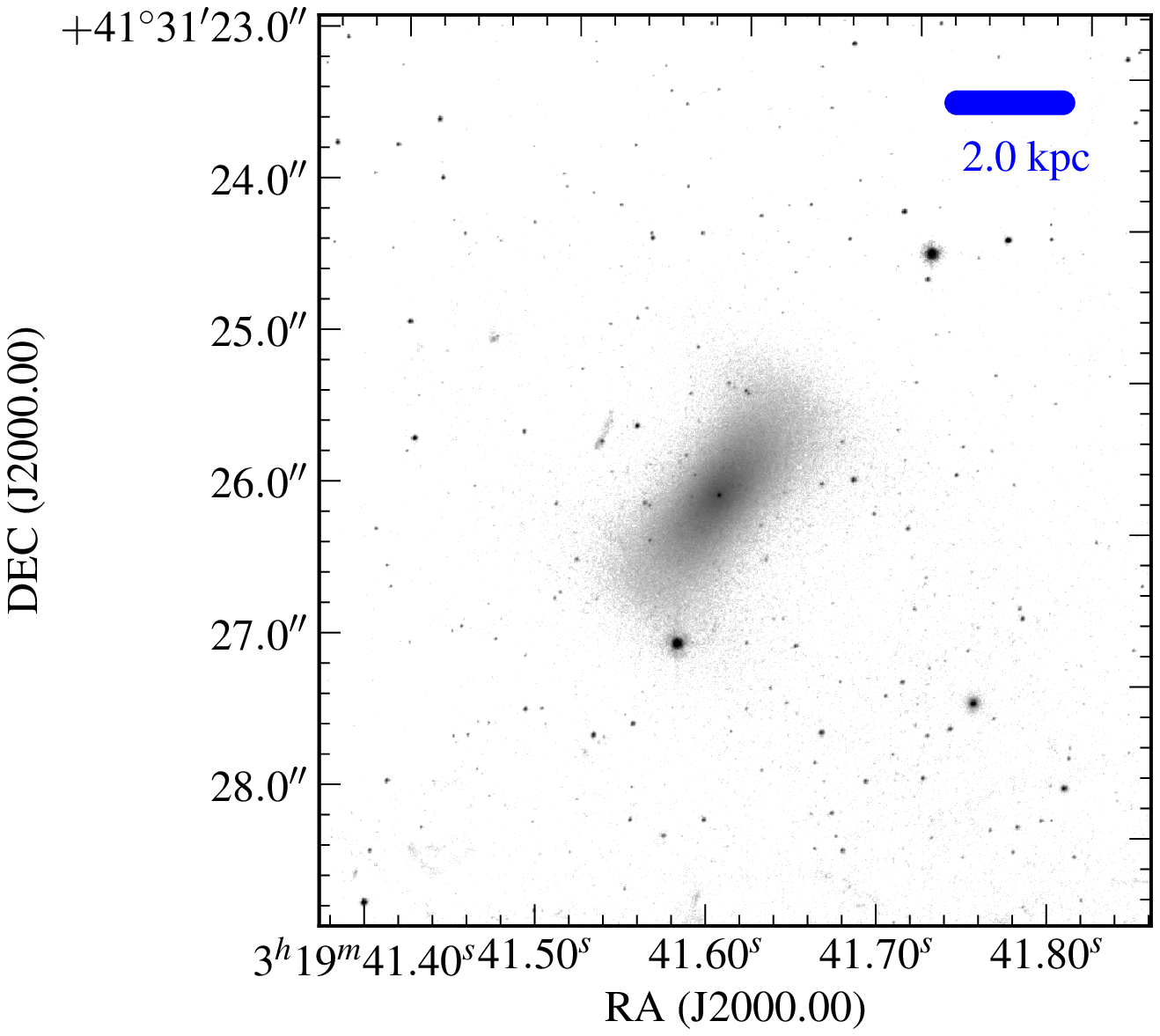}\includegraphics[width=0.47\textwidth]{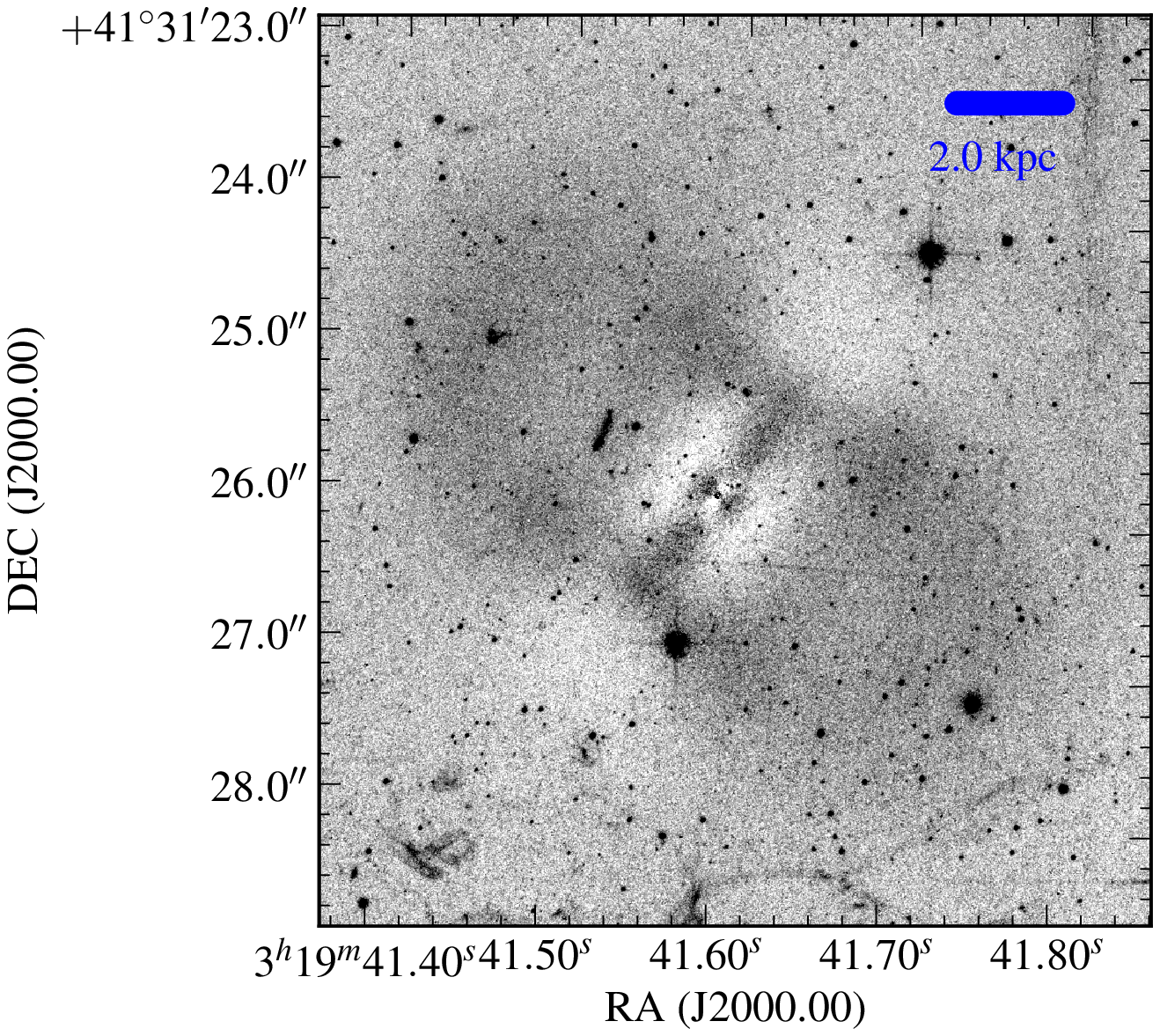}
\caption{HST-ACS imaging of the Perseus Cluster dwarf elliptical SA~0426-002. The left-hand panel shows the dE with the background light from nearby NGC~1275 subtracted. The right-hand panel shows low surface brightness in the outer regions of the dE after the subtraction of the \textsc{galfit} model described in Section~\ref{sec:sg}. A pair of low surface brightness lobes are clearly seen. These wings/lobes are likely a ring feature, which we hypothesise is the result of a tidal interaction with either a massive cluster galaxy or the cluster tidal potential.\label{dwarf}}
\end{figure*}

To examine the morphology and scaling relations of SA~0426-002, we utilise high resolution archival \textit{Hubble Space Telescope} (HST) Advanced Camera for Surveys (ACS) Wide Field Channel imaging (Figure \ref{dwarf}). The imaging was obtained in the \textit{F625W} ($\sim R$) band, with a total exposure time of 2481~s, and a pixel scale $0.05''$.  The left-hand panel of Figure~\ref{dwarf} highlights features in the centre of the dwarf, including a bar-like morphology, and the model-subtracted image in the right-hand panel shows the low surface brightness wings/lobes exhibited by the dwarf. We utilise this high-resolution imaging to quantify the morphology of SA~0426-002 via an analysis of its light distribution. 

\section{Results}
\label{sec:res}

\subsection{The size-magnitude relation}
\label{sec:sg}

To better understand the nature of SA~0426-002 compared to ``normal'' dEs such as CGW38 and CGW39, we examine its scaling relations, including its location on the size-magnitude and $\sigma$-luminosity relations for early-type galaxies. First, we must determine the size of the dwarf. To do this, SA~0426-002 is modelled using \textsc{galfit} \citep{2002AJ....124..266P}.

Given its proximity to the brightest cluster galaxy (BCG) NGC~1275, it was necessary to first remove the BCG's light to allow SA~0426-002 to be modelled accurately. We model NGC~1275 using a S\'ersic model with a fixed S\'ersic index of $n=4$ and effective radius $R_{e} = 15$~kpc \citep{2008A&A...483..793S}, then subtract this model from the original image to remove the background gradient. SA~0426-002  was then modelled using a single profile S\'ersic model convolved with the \textit{HST} ACS point spread function (PSF). The best fit model has S\'ersic~$n=1.24\pm 0.02$ and $R_{e} = 2.1 \pm 0.10$~kpc. The model of SA~0426-002 is elongated, with an axis ratio $b/a = 0.47 \pm 0.1$ ($\epsilon=0.63$), and a boxy structure.  The residuals of this model fit are shown in the right-hand panel of Figure~\ref{dwarf}, revealing the wings/lobes that surround the dwarf. 

\begin{figure}
\includegraphics[width=0.47\textwidth]{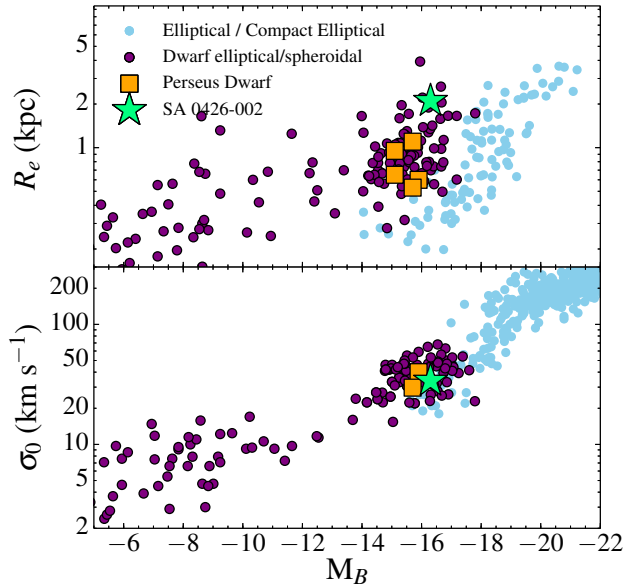}
\caption{The size-magnitude relation (top panel) and $\sigma$-luminosity relation (bottom panel) for pressure supported systems with $R_{e} > 150$~pc. Dwarf elliptical and dSph galaxies are plotted as purple dots to separate them from giant Es and cEs.  We plot Perseus dEs with redshifts measured in this work as orange squares, and highlight the unusual dwarf SA~0426-002 as a green star. All the Perseus dEs lie on the size-magnitude and $\sigma$-luminosity relations for early-type galaxies. \label{sizemag}}
\end{figure}

We plot the newly calculated size of  SA~0426-002 on the size-magnitude relation in Fig.~\ref{sizemag}, along with those of the Perseus dEs presented in this work, and a comparison sample of distance-confirmed early-type galaxies taken from the compilation of  \citet{2011AJ....142..199B}\footnote{Available for download at sages.ucolick.org/spectral\_database.html.} and references therein, with an update to this sample provided in \citet{2013MNRAS.435L...6F}  and references therein. The $V$-band magnitudes from Brodie et al. (2011) and Forbes et al. (2013) are converted to the $B$~band using $(B-V) = 0.96$, a typical colour for an elliptical galaxy  \citep{1995PASP..107..945F}. The sizes of the Perseus dwarfs in Fig.~\ref{sizemag} are taken from Conselice, Gallagher \& Wyse (2003) and \citet{2009MNRAS.393..798D}, with the exception of SA~0426-002, which we calculate above. 

The Brodie et al. (2011) sample contains a mix of compact ellipticals (cEs), dEs, and giant ellipticals (E), and the scaling relations of dEs and cE/E galaxies diverge at $M_{B} \approx -18$ \citep{1985ApJ...295...73K}. To separate the galaxy populations, we highlight all galaxies classed as dE or dSph as purple circles. The sizes of the spectroscopically confirmed Perseus Cluster members agree with those of dEs in other studies, though we note SA~0426-002 is one of the largest dwarfs identified to date at its luminosity ($M_{B} = -16.3$). We will discuss a possible reason for this extended nature in Section~\ref{sec:tidal}.  

\subsection{Surface photometry}
  
\begin{figure}
\includegraphics[width=0.47\textwidth]{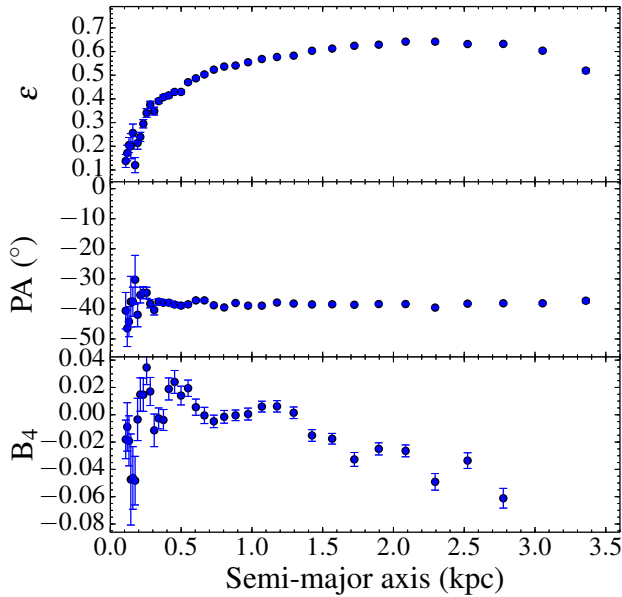}
\caption{Light profile fits to SA~0426-002, showing the ellipticity (top panel), position angle (middle panel), and 4th harmonic amplitude $B4$ (bottom panel) as a function of semi-major axis. The dwarf becomes increasingly boxy ($B_{4} <0$) as the semi-major axis increases. Combined with elongated isophotes, this is consistent with its bar-like morphology. The position angle is constant with semi-major axis, and therefore the dwarf does not exhibit twists in its isophotes.\label{surface}}
\end{figure}

To examine the light distribution of SA~0426-002 in greater detail, the dwarf was modelled using the \textsc{iraf} package \textsc{ellipse} to fit isophotes to its light distribution. The position angle (PA), ellipticity ($\epsilon$), and centre of the isophotes were allowed to vary during the fit. We show the results of this isophotal fit in Figure~\ref{surface}, excluding the inner 0.3'' of the profile fit where the effect of the ACS PSF is strongest. The ellipticity ($\epsilon$, top panel), position angle (PA, middle panel), and 4th harmonic amplitude $B4$ (bottom panel) are plotted as a function of semi-major axis to highlight the bar-like nature of the dwarf. The 4th harmonic amplitude $B4$ provides the deviation of the isophote from a perfect ellipse, with $B4=0$ for a perfect ellipse. Discy isophotes will have positive values of $B4$, while boxy isophotes will have negative values.  

The position angle of the isophotal fit remains constant at PA~$\approx-40^{\circ}$ out to the last plotted isophote at $R~=~3.5$~kpc, with no evidence for warps//twists in the isophotes. In contrast, the dwarf's ellipticity changes from near circular in the centre of the galaxy, to elongated isophotes with $\epsilon=0.63$ at $R=2.5$~kpc. The diskyness/boxiness of the dwarf furthermore changes dramatically over the structure of the bar feature, with the isophote's 4th harmonic amplitude $B4 \sim 0.02$ at $R=0.5$~kpc (i.e. discy isophotes), compared to boxy isophotes with $B4 \sim -0.03$ at $R=2.5$~kpc. These profiles are highlight the dwarfs bar-like structure. These results are consistent with the surface photometry presented in Conselice et al. (2002).

\subsection{Central velocity dispersions and dynamical masses}
\label{sec:veldisps}

If a dE is formed via the morphological transformation and tidal stripping of a disc galaxy, we might expect it to have an unusually high velocity dispersion for its luminosity as $\sigma_{0}$ is expected to be relatively unchanged by the interaction \citep{1992ApJ...399..462B}, whereas its luminosity will be reduced. This will result in a high dynamical mass-to-light ratio for the dwarf. We therefore use our \textsc{esi} spectroscopy to measure central velocity dispersions and dynamical masses for three dwarfs in our sample with a minimum signal-to-noise of 7 at $\sim8500$~\AA: SA~0426-002, CGW38, and CGW39. 

To  measure global internal velocity dispersions $\sigma$ for the dEs, we use the Penalised Pixel Fitting (\textsc{Ppxf}) routine \citep{2004PASP..116..138C}. We utilise five standard stars taken under an identical instrumental setup to the dwarf observations.   The Fe, Mg, CaT, and H$\alpha$ lines are used in determining the velocity dispersion for CGW39, and the H$\alpha$ line used for CGW38 and SA~0426-002 due to the CaT feature blending with skylines.   For SA~0426-002, we measure $\sigma = 32.4 \pm 5.5$~km~s$^{-1}$. We measure $\sigma = 38.8 \pm 4.4$~km~s$^{-1}$ for CGW38, and $\sigma = 26.3\pm4.3$~km~s$^{-1}$ for CGW39. The errors are the formal errors provided by \textsc{ppxf}.

To compare our results to the literature, we convert these global values of $\sigma$ to central values $\sigma_{0}$ at $R_{e}/8$ using equation 1 of \citet{2006MNRAS.366.1126C}. These conversions are $<10\%$ of the aperture velocity dispersion, and within the error bars of the uncorrected values of $\sigma$.  For SA~0426-002, we find $\sigma_{0} = 33.9 \pm 6.1$~km~s$^{-1}$. CGW38 has a central velocity dispersion $\sigma_{0} = 40.0 \pm 5.5$~km~s$^{-1}$, and for CGW39, we determine a central velocity dispersion $\sigma_{0} = 29.7 \pm 7.3$~km~s$^{-1}$.  We plot these central velocity dispersions in Fig.~\ref{sizemag}, along with a comparison sample taken from \citet{2014MNRAS.439.3808P}. All three Perseus dEs lie on the $\sigma$-luminosity relation for early-type galaxies, and do not have unusual values of $\sigma_{0}$ for their magnitudes. 

Using these central velocity dispersions, we calculate dynamical masses $M_{dyn}$ for the three dEs using the expression

\begin{equation}
M_{dyn} = C\sigma^{2}R,
\end{equation}

\noindent where $R$ is the size of the galaxy, $\sigma$ the object's velocity dispersion, and $C$ is a correction for the the object's light profile. For an exponential S\'ersic brightness profile with $n=1.0$ (typical for a dwarf elliptical galaxy), $C\approx8$ \citep{2002A&A...386..149B} provides the total mass of the galaxy. 

For SA~0426-002, this provides a dynamical mass $M_{dyn} = (4.5\pm2.3)\times10^{9}$~M$_{\odot}$.  However, we note that given its strong bar-like structure, SA~0426-002 is likely rotating, and the conversion of the central velocity dispersion to a dynamical mass is therefore approximate. CGW38 has $M_{dyn} = (1.8 \pm 0.9)\times10^{9}$~M$_{\odot}$, and CGW39 has $M_{dyn} = (8.9\pm 0.5)\times10^{8}$~M$_{\odot}$. These dynamical masses, along with the central velocity dispersions $\sigma_{0}$, are shown in Table~\ref{dwarfprops}. 

\subsection{Stellar masses and mass-to-light ratios}

To investigate if any of the dwarfs have unusually high dynamical mass to stellar mass ratios (as might be expected for a tidally stripped object), we calculate stellar masses for all three dwarfs. This requires an estimation of both the ages and metallicities of the dwarfs stellar populations, as the stellar mass-to-light ratio is dependent on both factors.  Penny \& Conselice (2008) estimate CGW38 and CGW39 to have old ($>10$~Gyr) stellar populations, with metallicities [Z/H]~$<-0.33$~dex. To calculate stellar mass-to-light ratios, we utilise the SSP models of Bruzual \& Charlot (2003), assuming a Salpeter IMF,  a stellar age of 10~Gyr, and a metallicity [Fe/H]$~=~-0.33$~dex.
 
Based on these assumptions,  CGW38 and CGW39 have $B$-band stellar mass-to-light ratios $\sim6$. We note this ratio is an upper limit, as lower metallicity stellar populations will have lower stellar mass-to-light ratios. By converting their luminosities to stellar masses, we find $M_{\star} = (1.9 \pm 0.3)\times10^{9}$~M$_{\odot}$ for CGW38, and $M_{\star}= (1.7 \pm 0.5) \times 10^{9}$~M$_{\odot}$ for CGW39. Errors are calculated based on the uncertainty in the ages and metallicities of the dwarfs stellar populations. Within the error bars, these stellar masses are consistent with their dynamical masses. 

We repeat this stellar mass calculation for SA~0426-002. However, there is no estimation of the metallicity or age of this galaxy's stellar population. We therefore compare its colour to Bruzual \& Charlot (2003) models to obtain the age of its stellar population, assuming an upper limit on its metallicity of [Fe/H]~$=-0.33$~dex. SA~0426-002 has a colour $(B-R) = 1.3$ (Conselice et al., 2002). Comparing this colour to the  Bruzual \& Charlot (2003) stellar population synthesis models, the dwarf  ceased star formation within the last 5~Gyr, and must be older than 1~Gyr.

Using the models of Bruzual \& Charlot (2003), assuming a Salpeter IMF, an upper age limit 5~Gyr, and metallicity [Fe/H]~$\sim -0.33$, we estimate SA~0426-002 to have a maximum $B$-band stellar M/L ratio $\sim3.6$. This gives a stellar mass $(1.9\pm 0.8)\times 10^{9}$~M$_{\odot}$.  Its stellar mass is lower than its dynamical mass of $(4.5\pm1.6)\times10^{9}$~M$_{\odot}$, giving a dynamical mass to stellar mass ratio of $\sim2$. This low mass-to-light ratio suggests the dwarf is unlikely to be dark matter dominated in its inner regions, consistent with the results of Penny et al. (2009) for the mass-to-light ratios of bright dwarf ellipticals in Perseus. The stellar masses of all three dwarfs are listed in Table~\ref{dwarfprops}, along with errors based on uncertainties in the ages of the dwarfs.

\section{Discussion}
\label{sec:discuss}

All three Perseus dwarfs with measured kinematics have typical central velocity dispersions for their luminosities (Fig.~2), and lie on the size-magnitude relation for early-type galaxies, though SA~0426-002 is one of the most extended dwarfs at its luminosity. Given its unusual morphology (Fig.~\ref{dwarf}), we discuss this dwarf in greater detail. 

SA~0426-002 is remarkably symmetric in appearance, with a bar-like morphology extending to $R\approx2.8$~kpc (8'') from its nucleus. The central regions of the galaxy resemble an elongated dwarf elliptical with a boxy morphology. The galaxy has a nuclear star cluster, typical for bright dwarf ellipticals \citep{2006ApJS..165...57C}. However, the outer regions of this dwarf elliptical are highly unusual. Two symmetric, low surface brightness ``wings'' can be seen either side of the main body of the galaxy (Fig.~\ref{dwarf}).  Conselice, Gallagher and Wyse (2002) measure these wings to have $\mu_{B} = 27$ mag arcsec$^{-2}$. We discuss a possible origin for this unusual structure. 

\subsection{A tidally harassed galaxy?}
\label{sec:tidal}

SA 0426-002 is a unique dwarf galaxy, and the authors are unable to locate a dE with comparable morphology to SA 0426-002 in the literature. While bar-like dEs are known to exist in the Virgo Cluster \citep{2012ApJ...750..121G}, these dwarfs do not exhibit the low-surface brightness lobes seen for SA~0426-002 in SDSS imaging of comparable depth.  

One possible explanation for this unusual morphology is that SA~0426-002 is being morphologically transformed from a low mass disc or dwarf irregular to a dwarf elliptical via harassment. Due to  multiple high-speed interactions with other cluster members, the infalling disc has been tidally perturbed, triggering bar formation. During this process, stars in the outer region of the disc are tidally stripped, forming arms/tidal features (e.g. Mastropietro et al., 2005), resulting in a high effective radius $R_{e}$. for the galaxy's luminosity  Eventually, only the central bar will remain. We therefore hypothesise that SA 0426-002 has not yet had its disc completely stripped by high-speed interactions, and the low surface brightness wings are stars being tidally stripped from the galaxy. 

Simulations of bar formation in tidally transforming discs have focused heavily on the Milky Way satellite system (e.g. \citealt{2011ApJ...726...98K,2014arXiv1404.1211L}). While NGC~1275 is more massive than the Milky Way, the tidal transformation of low mass discs by its gravitational potential will progress in a similar, though more rapid, fashion, with bars forming on timescales $\sim1$~Gyr \citep{2005MNRAS.364..607M,2009A&A...494..891A} . The simulations of Mayer et al. (2001) show that a rotationally supported, discy dIrr infalling  into a MW-sized halo will form a bar with ring-like structures comparable to the morphology of SA~0426-002 . Bar formation is  rapid, occurring after the first pericentric approach ({\L}okas et al., 2014).  The morphology of the dwarf is also remarkably similar to the simulations of Mastropietro et al. (2005), \citet{2009A&A...494..891A}, and \citet{2012MNRAS.424.2401V}, particularly for face-on views of simulated tidally interacting dwarfs. SA~0426-002 is likely a rare example of a dwarf midway through the morphological transformation from a disc to dE, and we are viewing this transformation face-on. 
 
\citet{2009A&A...494..891A} present simulations of low mass disc galaxies with varying bulge/disc ratios undergoing harassment in the cluster environment. The morphology of SA~0426-002 is remarkably similar to their simulated dwarfs whose progenitors have small bulge components and prograde rotation. In their simulations, \citet{2009A&A...494..891A} show that the scale-length of their discs will be reduced by 40-50\% after several tidal interactions, and the remnants will lie on the fundamental plane for dwarf ellipticals. The size-luminosity relation for dwarf ellipticals is consistent with that of a population of truncated disc galaxies \citep{2008A&A...489.1015B}, further strengthening the argument that at least a fraction of dEs are tidally stripped disc galaxies. However, SA~0426-002 is extended for its luminosity on the size-magnitude relation presented in Fig.~\ref{sizemag}, and appears not to fit this scenario. 

The fact that SA~0426-002 appears extended on the size-magnitude relation can be explained as we are likely seeing the dwarf early in its transformation from a disc to dwarf elliptical via harassment. The intra-cluster medium will rapidly ram-pressure strip a dwarf of its gas, and the dwarf will move to a fainter location on the size-magnitude relation, but it will have an extended size for its luminosity.   SA~0426-002's colour $(B-R)=1.3$ \citep{2002AJ....123.2246C} places it on the red-sequence for dwarf ellipticals presented in \citet{2008MNRAS.383..247P}, and its \textsc{keck-esi} spectrum does not exhibit any emission lines consistent with ongoing star formation (e.g. H$\alpha$). This lack of star formation and red colour shows the dwarf has already been stripped of gas by the cluster environment. However, further tidal interactions are required to strip its mass, and hence decrease its effective radius. The ring feature in Fig.~\ref{dwarf} can be easily destroyed by gravitational heating due to further tidal interactions in the cluster environment \citep{2005MNRAS.364..607M}, which will reduce the dwarf's effective radius.

Given its projected separation of 35~kpc from the brightest cluster galaxy NGC~1275, could SA~0426-002 and the BCG be interacting? NGC~1275 is located at $v= 5276$~km~s$^{-1}$ (Petrosian et al., 2007), with SA~0426-002 at $7987$~km~s$^{-1}$. With a velocity separation of $2711$~km~s$^{-1}$, the two galaxies are either undergoing a high-speed interaction, else SA~0426-002 is infalling into the cluster and interacting with the cluster tidal potential. \citet{2009ApJ...706L.124L} show that dwarfs with flattened morphologies (similar to SA~0426-002) in the Virgo Cluster typically have fast line-of-sight velocities, whereas spherical dEs are typically slow. In this scenario, this difference in morphology and velocity distributions for round vs. flattened dwarfs is taken as evidence for an infall population with eccentric orbits. Flattened dwarfs represent the recent cluster infall population, whereas round dwarfs are the ``first'' generation of cluster dEs which have may have formed inside the primordial cluster itself.   

\subsection{A ring galaxy?}
\label{ringgal}

We consider the likelihood that SA~0436-002 is a ring galaxy.  Examples of galaxies in the nearby Universe with comparable morphology to SA~0426-002 are ESO325-28, ESO509-98, and NGC~2665. However, these ring galaxies are massive, star forming systems brighter than $M_{B}=-19.5$. Furthermore, SA~0436-002 lacks the spiral arms seen in these ring galaxies.

Given its strong bar-like morphology, a bar resonance has likely  been set up in the galaxy at the outer Linblad resonance, resulting in the low surface brightness lobes either side of the central bar.  The dwarf exhibits the characteristic ``dimpled'' outer ring morphology, similar to the more inclined galaxy ESO~287-56 \citep{1995ApJS...96...39B}. Gravitational torques from the bar can depopulate material at the co-rotation radius while building it up at the outer Lindblad Radius. Thus both a bar and gas in a disk are required to form the lobes or wings seen in SA~0426-002.  We therefore hypothesise that the progenitor of SA~0426-002 was a dwarf disc galaxy, as gas is required to form the stars which make up the lobe features. 

This ring galaxy morphology does not exclude ongoing tidal interactions; indeed, such interactions can trigger bar formation in an infalling disc galaxy. Mastropietro et al. (2005) show that infalling discs will undergo major structural transformation, including strong bar formation and spiral patterns. These spiral features will be stripped away, with remaining material forming a ring structure around the bar. Several galaxies in their simulations exhibit bar and ring features, and thus  SA~0426-002 has characteristics of a ring galaxy formed via tidal harassment. Further tidal interactions within Perseus will likely remove the remaining ring structure, until only the bar of the galaxy remains, and such dwarfs are not uncommon in clusters (e.g \citealt{2012ApJ...750..121G}).

\section{Conclusions}
\label{sec:conc}

Using the Keck-\textsc{ESI} spectrograph, we confirm two dEs  as a member of the Perseus Cluster (CGW40 and SA~0426-002), and re-confirm cluster membership for four other dwarfs. The dwarfs lie on the size-magnitude relation for dEs, but SA~0426-002 has a large size for its magnitude, as it has ceased star formation but not yet been completely tidally stripped. We furthermore calculate central velocity dispersions and dynamical masses for three dwarfs in our sample: CGW38, CGW39, and SA~0426-002, and all three dwarfs follow the $\sigma$-luminosity relation for early-type stellar systems.  

We examine the unusual dwarf SA~0426-002 in detail. SA~0426-002 exhibits a bar-like morphology with $R_{e} = 2.1\pm0.1$~kpc and S\'ersic~$n = 1.24\pm0.02$. Based on its bar-like morphology, ring structure, and its location in the Perseus Cluster, we hypothesise that SA~0426-002 is near its orbit pericentre, and is currently undergoing morphological transformation from a disc galaxy into a dwarf elliptical. The wing-like structure is the result of stellar mass loss due to tidal heating by repeated interactions with other cluster galaxies, else tidal interactions with the BCG NGC~1275. 

Combined with embedded substructure identified in deep imaging of nearby dwarf ellipticals in Virgo (e.g. \citealt{2009A&A...501..429L}) and Fornax (e.g. \citealt{2003A&A...400..119D}), it is becoming increasingly apparent that a fraction of dwarf ellipticals are not part of the primordial cluster population. Instead, these objects are formed at later times via the transformation of infalling disc galaxies. 

\section*{Acknowledgments}

SJP and KAP acknowledge the support of an Australian Research Council Super Science Fellowship grant FS110200047. DAF thanks the ARC for support via DP130100388. We thank Michael JI Brown and Alister Graham for useful discussions. We thank the referee for their thorough review which has improved the paper. The data presented herein were obtained at the W.M. Keck Observatory, which is operated as a scientific partnership among the California Institute of Technology, the University of California and the National Aeronautics and Space Administration. The Observatory was made possible by the generous financial support of the W.M. Keck Foundation. The authors wish to recognise and acknowledge the very significant cultural role and reverence that the summit of Mauna Kea has always had within the indigenous Hawaiian community, and we are most fortunate to have the opportunity to conduct observations from this mountain.

\bsp

\label{lastpage}


\begin{thebibliography}{}
\bibitem[\protect\citeauthoryear{Aguerri 
\& Gonz{\'a}lez-Garc{\'{\i}}a}{2009}]{2009A&A...494..891A} Aguerri J.~A.~L., Gonz{\'a}lez-Garc{\'{\i}}a A.~C., 2009, A\&A, 494, 891 
\bibitem[\protect\citeauthoryear{Bender, Burstein, 
\& Faber}{1992}]{1992ApJ...399..462B} Bender R., Burstein D., Faber S.~M., 1992, ApJ, 399, 462
\bibitem[\protect\citeauthoryear{Bertin, Ciotti, 
\& Del Principe}{2002}]{2002A&A...386..149B} Bertin G., Ciotti L., Del Principe M., 2002, A\&A, 386, 149
\bibitem[\protect\citeauthoryear{Boselli et 
al.}{2008}]{2008A&A...489.1015B} Boselli A., Boissier S., Cortese L., Gavazzi G., 2008, A\&A, 489, 1015 
\bibitem[\protect\citeauthoryear{Brodie et al.}{2011}]{2011AJ....142..199B} 
Brodie J.~P., Romanowsky A.~J., Strader J., Forbes D.~A., 2011, AJ, 142, 
199 
\bibitem[\protect\citeauthoryear{Bruzual 
\& Charlot}{2003}]{2003MNRAS.344.1000B} Bruzual G., Charlot S., 2003, MNRAS, 344, 1000
\bibitem[\protect\citeauthoryear{Buta}{1995}]{1995ApJS...96...39B} Buta R., 
1995, ApJS, 96, 39
\bibitem[\protect\citeauthoryear{Cappellari 
\& Emsellem}{2004}]{2004PASP..116..138C} Cappellari M., Emsellem E., 2004, PASP, 116, 138
\bibitem[\protect\citeauthoryear{Cappellari et 
al.}{2006}]{2006MNRAS.366.1126C} Cappellari M., et al., 2006, MNRAS, 366, 
1126
\bibitem[\protect\citeauthoryear{Conselice, Gallagher, 
\& Wyse}{2002}]{2002AJ....123.2246C} Conselice C.~J., Gallagher J.~S., III, Wyse R.~F.~G., 2002, AJ, 123, 2246
\bibitem[\protect\citeauthoryear{Conselice, Gallagher, 
\& Wyse}{2003}]{2003AJ....125...66C} Conselice C.~J., Gallagher J.~S., III, Wyse R.~F.~G., 2003, AJ, 125, 66 
\bibitem[\protect\citeauthoryear{C{\^o}t{\'e} et 
al.}{2006}]{2006ApJS..165...57C} C{\^o}t{\'e} P., et al., 2006, ApJS, 165, 
57
\bibitem[\protect\citeauthoryear{De Rijcke et 
al.}{2003}]{2003A&A...400..119D} De Rijcke S., Dejonghe H., Zeilinger W.~W., Hau G.~K.~T., 2003, A\&A, 400, 119
\bibitem[\protect\citeauthoryear{de Rijcke et 
al.}{2009}]{2009MNRAS.393..798D} de Rijcke S., Penny S.~J., Conselice 
C.~J., Valcke S., Held E.~V., 2009, MNRAS, 393, 798
\bibitem[\protect\citeauthoryear{Forbes et al.}{2013}]{2013MNRAS.435L...6F} 
Forbes D.~A., Pota V., Usher C., Strader J., Romanowsky A.~J., Brodie 
J.~P., Arnold J.~A., Spitler L.~R., 2013, MNRAS, 435, L6 
\bibitem[\protect\citeauthoryear{Fukugita, Shimasaku, 
\& Ichikawa}{1995}]{1995PASP..107..945F} Fukugita M., Shimasaku K., Ichikawa T., 1995, PASP, 107, 945
\bibitem[\protect\citeauthoryear{Graham, Jerjen, 
\& Guzm{\'a}n}{2003}]{2003AJ....126.1787G} Graham A.~W., Jerjen H., Guzm{\'a}n R., 2003, AJ, 126, 1787 
\bibitem[\protect\citeauthoryear{Graham et al.}{2012}]{2012ApJ...750..121G} 
Graham A.~W., Spitler L.~R., Forbes D.~A., Lisker T., Moore B., Janz J., 
2012, ApJ, 750, 121
\bibitem[\protect\citeauthoryear{Kazantzidis et 
al.}{2011}]{2011ApJ...726...98K} Kazantzidis S., {\L}okas E.~L., Callegari 
S., Mayer L., Moustakas L.~A., 2011, ApJ, 726, 98
\bibitem[\protect\citeauthoryear{Kormendy}{1985}]{1985ApJ...295...73K} 
Kormendy J., 1985, ApJ, 295, 73 
\bibitem[\protect\citeauthoryear{Jerjen, Kalnajs, 
\& Binggeli}{2000}]{2000A&A...358..845J} Jerjen H., Kalnajs A., Binggeli B., 2000, A\&A, 358, 845
\bibitem[\protect\citeauthoryear{Lisker, Grebel, 
\& Binggeli}{2006}]{2006AJ....132..497L} Lisker T., Grebel E.~K., Binggeli B., 2006, AJ, 132, 497
\bibitem[\protect\citeauthoryear{Lisker 
\& Fuchs}{2009}]{2009A&A...501..429L} Lisker T., Fuchs B., 2009, A\&A, 501, 429
\bibitem[\protect\citeauthoryear{Lisker et al.}{2009}]{2009ApJ...706L.124L} 
Lisker T., et al., 2009, ApJ, 706, L124 
\bibitem[\protect\citeauthoryear{Lokas et al.}{2014}]{2014arXiv1404.1211L} 
Lokas E.~L., Athanassoula E., Debattista V.~P., Valluri M., del Pino A., 
Semczuk M., Gajda G., Kowalczyk K., 2014, arXiv, arXiv:1404.1211
\bibitem[\protect\citeauthoryear{Mastropietro et 
al.}{2005}]{2005MNRAS.364..607M} Mastropietro C., Moore B., Mayer L., 
Debattista V.~P., Piffaretti R., Stadel J., 2005, MNRAS, 364, 607
\bibitem[\protect\citeauthoryear{Mayer et al.}{2001}]{2001ApJ...559..754M} 
Mayer L., Governato F., Colpi M., Moore B., Quinn T., Wadsley J., Stadel 
J., Lake G., 2001, ApJ, 559, 754 
\bibitem[\protect\citeauthoryear{Mayer et al.}{2007}]{2007Natur.445..738M} 
Mayer L., Kazantzidis S., Mastropietro C., Wadsley J., 2007, Natur, 445, 
738
\bibitem[\protect\citeauthoryear{Moore et al.}{1996}]{1996Natur.379..613M} 
Moore B., Katz N., Lake G., Dressler A., Oemler A., 1996, Natur, 379, 613
\bibitem[\protect\citeauthoryear{Moore, Lake, 
\& Katz}{1998}]{1998ApJ...495..139M} Moore B., Lake G., Katz N., 1998, ApJ, 495, 139
\bibitem[\protect\citeauthoryear{Peng et al.}{2002}]{2002AJ....124..266P} 
Peng C.~Y., Ho L.~C., Impey C.~D., Rix H.-W., 2002, AJ, 124, 266 
\bibitem[\protect\citeauthoryear{Penny 
\& Conselice}{2008}]{2008MNRAS.383..247P} Penny S.~J., Conselice C.~J., 2008, MNRAS, 383, 247 
\bibitem[\protect\citeauthoryear{Penny et al.}{2009}]{2009MNRAS.393.1054P} 
Penny S.~J., Conselice C.~J., de Rijcke S., Held E.~V., 2009, MNRAS, 393, 
1054
\bibitem[\protect\citeauthoryear{Penny et al.}{2014}]{2014MNRAS.439.3808P} 
Penny S.~J., Forbes D.~A., Strader J., Usher C., Brodie J.~P., Romanowsky 
A.~J., 2014, MNRAS, 439, 3808 
\bibitem[\protect\citeauthoryear{Ry{\'s}, van de Ven, 
\& Falc{\'o}n-Barroso}{2014}]{2014MNRAS.439..284R} Ry{\'s} A., van de Ven G., Falc{\'o}n-Barroso J., 2014, MNRAS, 439, 284
\bibitem[\protect\citeauthoryear{Salom{\'e} et 
al.}{2008}]{2008A&A...483..793S} Salom{\'e} P., Revaz Y., Combes F., Pety J., Downes D., Edge A.~C., Fabian A.~C., 2008, A\&A, 483, 793 
\bibitem[\protect\citeauthoryear{Sheinis et 
al.}{2002}]{2002PASP..114..851S} Sheinis A.~I., Bolte M., Epps H.~W., 
Kibrick R.~I., Miller J.~S., Radovan M.~V., Bigelow B.~C., Sutin B.~M., 
2002, PASP, 114, 851 
\bibitem[\protect\citeauthoryear{Toloba et 
al.}{2011}]{2011A&A...526A.114T} Toloba E., Boselli A., Cenarro A.~J., Peletier R.~F., Gorgas J., Gil de Paz A., Mu{\~n}oz-Mateos J.~C., 2011, A\&A, 526, A114
\bibitem[\protect\citeauthoryear{Villalobos et 
al.}{2012}]{2012MNRAS.424.2401V} Villalobos {\'A}., ., De Lucia G., Borgani 
S., Murante G., 2012, MNRAS, 424, 2401
\end{thebibliography}
\end{document}